\documentclass[11pt,a4paper]{article}
\pdfoutput=1 

\RequirePackage{ifpdf} 
\usepackage{amsmath} 
\usepackage{mathtools}

\usepackage{jheppub}
\usepackage{pstricks}
\usepackage[final]{pdfpages} 
\usepackage{ifpdf} 
\usepackage{slashed}

\usepackage{color,xcolor}
\definecolor{urlblue}{rgb}{0.2,0.4,0.7}
\definecolor{citegreen}{rgb}{0,0.6,0.2}
\definecolor{linkred}{rgb}{0.9,0.2,0.1}
\usepackage{hyperref}
\hypersetup{
colorlinks=true, citecolor=citegreen, linkcolor=blue, urlcolor=urlblue}

\usepackage{graphics}
\usepackage{etoolbox} 
\usepackage{fixmath}
\usepackage{psfrag}

\usepackage{notoccite} 

\usepackage{amsfonts}
\usepackage{autobreak}

\usepackage{tikz}
\usetikzlibrary{positioning,arrows}
\usetikzlibrary{decorations.pathmorphing}
\usetikzlibrary{decorations.markings}
\usetikzlibrary{shapes.geometric}
\tikzset{
    vector/.style={decorate, decoration={snake}, draw},
    provector/.style={decorate, decoration={snake,amplitude=2.5pt}, draw},
    antivector/.style={decorate, decoration={snake,amplitude=-2.5pt}, draw},
    fermion/.style={draw=black, postaction={decorate},decoration={markings,mark=at position .55 with {\arrow[draw=black]{>}}}},
    fermionbar/.style={draw=black, postaction={decorate},
                       decoration={markings,mark=at position .55 with {\arrow[draw=black]{<}}}},
    fermionnoarrow/.style={draw=black},
    gluon/.style={decorate, draw=black,decoration={coil,amplitude=4pt, segment length=5pt}},
    scalar/.style={dashed,draw=black, postaction={decorate},decoration={markings,mark=at position .55 with {\arrow[draw=black]{>}}}},
    scalarbar/.style={dashed,draw=black, postaction={decorate},decoration={markings,mark=at position .55 with {\arrow[draw=black]{<}}}},
    scalarnoarrow/.style={dashed,draw=black},
    electron/.style={draw=black, postaction={decorate},decoration={markings,mark=at position .55 with {\arrow[draw=black]{>}}}},
    bigvector/.style={decorate, decoration={snake,amplitude=4pt}, draw},
}

\title{Two-loop Doubly Massive Four-Point Amplitude Involving a half-BPS and Konishi Operator}

\author{Taushif Ahmed$^{a}$ and Prasanna K. Dhani$^{b}$}
\emailAdd{taushif@mpp.mpg.de,dhani@fi.infn.it}

\affiliation{$^a$Max-Planck-Institut f\"ur Physik, Werner-Heisenberg-Institut, 80805 M\"unchen, Germany \\
$^b$INFN, Sezione di Firenze, I-50019 Sesto Fiorentino, Florence, Italy}

\preprint{MPP-2019-20}

\abstract{The two-loop four-point amplitude of two massless SU(N) colored scalars and two color singlet operators with different virtuality described by a half-BPS and Konishi operators is calculated analytically in maximally supersymmetric Yang-Mills theory. We verify the ultraviolet behaviour of the unprotected composite operator and exponentiation of the infrared divergences with correct universal values of the anomalous dimensions in the modified dimensional reduction scheme. The amplitude is found to contain lower transcendental weight terms in addition to the highest ones and the latter has no similarity with similar amplitudes in QCD.}

\begin{document}
\allowdisplaybreaks[4]
\unitlength1cm
\maketitle
\flushbottom


\def\D{{\cal D}}
\def\DD{\overline{\cal D}}
\def\g{\overline{\cal G}}
\def\gm{\gamma}
\def\M{{\cal M}}
\def\ep{\epsilon}
\def\epm1{\frac{1}{\epsilon}}
\def\epm2{\frac{1}{\epsilon^{2}}}
\def\epm3{\frac{1}{\epsilon^{3}}}
\def\epm4{\frac{1}{\epsilon^{4}}}
\def\unM{\hat{\cal M}}
\def\ashat{\hat{a}_{s}}
\def\asmur{a_{s}^{2}(\mu_{R}^{2})}
\def\sigbar{{{\overline {\sigma}}}\left(a_{s}(\mu_{R}^{2}), L\left(\mu_{R}^{2}, m_{H}^{2}\right)\right)}
\def\sigbarn{{{{\overline \sigma}}_{n}\left(a_{s}(\mu_{R}^{2}) L\left(\mu_{R}^{2}, m_{H}^{2}\right)\right)}}
\def\unas{ \left( \frac{\hat{a}_s}{\mu_0^{\epsilon}} S_{\epsilon} \right) }
\def\rnM{{\cal M}}
\def\bt{\beta}
\def\cD{{\cal D}}
\def\cC{{\cal C}}
\def\ca{\text{\tiny C}_\text{\tiny A}}
\def\cf{\text{\tiny C}_\text{\tiny F}}
\def\ct{{\red []}}
\def\sv{\text{SV}}
\def\murOmu{\left( \frac{\mu_{R}^{2}}{\mu^{2}} \right)}
\def\bb{b{\bar{b}}}
\def\bt0{\beta_{0}}
\def\bt1{\beta_{1}}
\def\bt2{\beta_{2}}
\def\bt3{\beta_{3}}
\def\gm0{\gamma_{0}}
\def\gm1{\gamma_{1}}
\def\gm2{\gamma_{2}}
\def\gm3{\gamma_{3}}
\def\nn{\nonumber}
\def\l{\left}
\def\r{\right}
\def\T{{\cal Z}}    
\def\U{{\cal Y}}
\def\qgraf{{\fontfamily{qcr}\selectfont
QGRAF}}
\def\python{{\fontfamily{qcr}\selectfont
PYTHON}}
\def\form{{\fontfamily{qcr}\selectfont
FORM}}
\def\reduze{{\fontfamily{qcr}\selectfont
REDUZE2}}
\def\kira{{\fontfamily{qcr}\selectfont
Kira}}
\def\litered{{\fontfamily{qcr}\selectfont
LiteRed}}
\def\fire{{\fontfamily{qcr}\selectfont
FIRE5}}
\def\air{{\fontfamily{qcr}\selectfont
AIR}}
\def\mint{{\fontfamily{qcr}\selectfont
Mint}}
\def\hepforge{{\fontfamily{qcr}\selectfont
HepForge}}
\def\arXiv{{\fontfamily{qcr}\selectfont
arXiv}}
\def\Python{{\fontfamily{qcr}\selectfont
Python}}
\def\anci{{\fontfamily{qcr}\selectfont
Finite\_ppbk.m}}

\newcommand{\dis}{}
\newcommand{\overbar}[1]{mkern-1.5mu\overline{\mkern-1.5mu#1\mkern-1.5mu}\mkern
1.5mu}


\section{Introduction}
\label{sec:intro}
Scattering amplitudes and correlation functions are the most fundamental objects in any quantum field theory (QFT). Gauge theory is the language of nature which is so far well tested through the standard model of particle physics. Even after many decades of the formal formulation of the Yang-Mills gauge theory~\cite{Yang:1954ek}, it remains a formidable task to go beyond a certain order in perturbation theory. In particular, very little is known in the non-perturbative regime. However, a special class of theories which admits a dual description in the strongly coupled sector are largely explored. The ${\cal N}=4$ supersymmetric Yang-Mills (SYM) belongs to this category. Through the AdS/CFT correspondence~\cite{Maldacena:1997re}, certain quantities in strongly coupled sectors of ${\cal N}=4$ SYM is related to the weakly coupled sectors of gravity in Anti-de Sitter space. Beyond the AdS/CFT correspondence, in particular, in the weakly coupled sector, the ${\cal N}=4$ SYM is seen to be a time tested sandbox to explore new ideas and computational techniques. It offers perhaps the best chance to solve an interacting four dimensional QFT. Due to the underlying superconformal symmetries, often it renders many computations much simpler than their non-supersymmetric counterparts. In many cases the resulting conceptual understanding and computational developments eventually help to solve problems in a generic QFT. 

Besides on-shell amplitudes, the study of form factors (FFs) has generated a surge of interest in the community. The FFs are a set of quantities which are constructed out of scattering amplitudes of on-shell states consisting of elementary particles of the theory and off-shell states described by composite operators. These are calculated by evaluating the quantity of the form $\langle p_1^{\sigma_1},\ldots, p_n^{\sigma_n}| {\cal O}|0 \rangle$ which represents the transition matrix element from vacuum $|0 \rangle$ to an on-shell state $|p_1^{\sigma_1},\ldots, p_n^{\sigma_n} \rangle$ through an interaction caused by the gauge invariant operator ${\cal O}$. $p_n$ and $\sigma_n$ represent the corresponding momentum and quantum numbers of the $n$-th particle. Studying these quantities is of paramount importance. For example, any $n$-point planar amplitude is factorised~\cite{Bern:2005iz} into an infrared divergent part described by a product of FF and a finite part, often called hard function. The divergent part exponentiates and the exponential term is described by some universal quantities like the light-like cusp anomalous dimension.

\begin{figure}
\begin{center}
\begin{tikzpicture}[line width=0.6 pt, scale=0.7]
\draw [dashed, blue] (-3,0) -- (0,-1.2);
\draw [dashed, blue] (-3,-3) -- (0,-1.8);
\draw [fill] (0.5,-1.5) circle [radius=0.5];
\draw [very thick] (0.7,-1.2) -- (4,0);
\draw [very thick] (0.7,-1.8) -- (4,-3);
\node at (-3.3,0.8) {${\rm m=0}$};
\node at (3.2,0.8) {${\rm m=m_1}$};
\node at (3.2,-3.8) {${\rm m=m_2}$};
\node at (-3.3,-3.8) {${\rm m=0}$};
\end{tikzpicture}
\caption{The four-point amplitude involving two color singlet states. The ${\rm m}$ represents the mass of the particle.}
\label{fig:scattering} 
\end{center}
\end{figure}
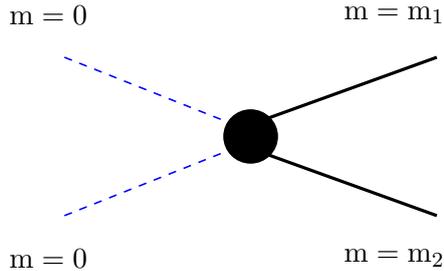

In recent times, several calculations~\cite{Brandhuber:2010ad,Bork:2010wf,Bork:2011cj,Brandhuber:2011tv,Gehrmann:2011xn,Brandhuber:2012vm,Boels:2015yna,Ahmed:2016vgl,Boels:2017ftb} have been carried out on FF in the context of ${\cal N}=4$ SYM. A very first calculation was done long back in~\cite{vanNeerven:1985ja} where the two-loop contribution to Sudakov FF of a half-BPS operator belonging to the stress-energy supermultiplet was performed. This was later extended to three loops in~\cite{Gehrmann:2011xn} where a remarkable connection to the highest transcendental terms of the corresponding FF in quantum chromodynamics (QCD) was found. There is another operator, called Konishi~\cite{KONISHI1984439}, which is the primary operator of the Konishi supermultiplet and belongs to the non-BPS category, has drawn a lot of attention due to many interesting properties it exhibits. It is the simplest gauge invariant operator in ${\cal N}=4$ SYM that is not protected by supersymmetry and consequently, receives non-zero anomalous dimensions to all orders in perturbation theory. The results of the anomalous dimensions up to five loops are known in the literature~\cite{Eden:2000mv,ANSELMI1997221,Bianchi:2000hn,Kotikov:2004er,Eden:2004ua,Bajnok:2008bm,Bajnok:2009vm,Fiamberti:2007rj,Fiamberti:2008sh,Velizhanin:2008jd,Eden:2012fe}. The two-point FF to two loops and three-point to one-loop were computed in~\cite{Nandan:2014oga} where the former was later extended by us in~\cite{Ahmed:2016vgl} to three loops and the latter one to two loops by one of us in~\cite{Banerjee:2016kri}. In this article, for the first time, we focus on a four-point amplitude of two different composite operators: half-BPS and Konishi. More specifically, we consider the four-point amplitude of two massless scalars (colored) in ${\cal N}=4$ SYM and two color singlet states with different masses represented through half-BPS and Konishi operators, see Fig.~\ref{fig:scattering}. In QCD, two loops four-point amplitudes involving two massive vector bosons $V_1V_2$ were computed in~\cite{Gehrmann:2015ora,vonManteuffel:2015msa}. The case for two identical operators (half-BPS/Konishi) is being investigated by us in~\cite{Ahmed:2019yjt}. For similar calculations in QCD involving di-Higgs boson, see~\cite{Banerjee:2018lfq,H:2018hqz}.

The multiloop corrections to the processes involving massive particles are known to be very complicated. Nevertheless, the present multiloop techniques seem to be able to stand the challenges of evaluation of the loop amplitudes. For the first time in ${\cal N}=4$ SYM, by employing state-of-the-art techniques we compute the four-point amplitude involving two massive particles with different virtuality to two loops level analytically. Unlike the most popular and relatively modern method of unitarity, we compute the amplitude by applying the Feynman diagrammatic approach. This approach is particularly useful in the context of regularisation prescription which is required in order to regulate the infrared (IR) divergences present in the theory due to the presence of massless particles. In addition, though the ${\cal N}=4$ SYM is ultraviolet (UV) finite in 4-dimensions, there can be UV divergences in the FF beyond leading order because of the composite operators. In~\cite{Nandan:2014oga}, it was shown that the FF of unprotected operators like Konishi calculated in four dimensional helicity (FDH) scheme~\cite{Bern:1991aq,Bern:2002zk} fail to produce the correct anomalous dimensions, instead in modified dimensional reduction (${\overline{\rm DR}}$)~\cite{Siegel:1979wq,Capper:1979ns} it indeed gives the correct results. Due to the similarity between the latter scheme with dimensional regularisation~\cite{tHooft:1972tcz} which is mostly used for the radiative corrections following Feynman diagrammatic approach, it is much more convenient to employ the ${\overline{\rm DR}}$ scheme. 

The degree of transcendentality, $\tau$, of a function $f$ is defined as the number of iterated integrals required to define the function $f$, e.g. $\tau(\log)=1\,, \tau({\rm Li}_n)=n\,, \tau(\zeta_n)=n$ and also we define $\tau(f_1f_2)=\tau(f_1)+\tau(f_2)$. Algebraic factors are assigned degree zero. It is an observed~\cite{Bern:2006ew,Drummond:2007cf,Naculich:2008ys,Bork:2010wf,Gehrmann:2011xn,Brandhuber:2012vm,Eden:2012rr,Drummond:2013nda,Basso:2015eqa,Goncalves:2016vir}, albeit unproven fact that the results of scattering amplitudes in ${\cal N}=4$ SYM exhibit uniform transcendentality (UT) i.e. those can be expressed in terms of polylogarithmic functions of uniform degree 2L, where L denotes the loop order, with constant coefficients. For the planar amplitudes, this is even true for individual integrals when these are expressed in an appropriate basis of dual conformal integrals~\cite{ArkaniHamed:2010gh,Drummond:2010mb}. However, for non-planar integrals the dual conformal symmetry does not hold true. For four points, the non-planar double ladder integral does not exhibit UT, however, if it is defined with an appropriate loop dependent numerator, it does obey the UT property~\cite{Drummond:2010mb,Drummond:2010cz}. So, by making an appropriate choice of basis, one can understand the UT property of four-point amplitudes in ${\cal N}=4$ SYM~\cite{Tausk:1999vh} and ${\cal N}=8$ supergravity amplitudes~\cite{Naculich:2008ew,Brandhuber:2008tf}. Inspired by these observations, a long-standing question is floating around: is the UT property a generic feature of ${\cal N}=4$ SYM amplitudes?  In~\cite{Nandan:2014oga,Ahmed:2016vgl}, it has been shown that UT property breaks down for the Sudakov FF of non-protected operators like Konishi. In this article, we address this question in the context of four-point amplitude involving two different color singlet states described by a half-BPS and Konishi. We see that the UT property does not hold true.

The paper is organised as follows. In Sec.~\ref{sec:th}, we introduce the Lagrangian, define the half-BPS and Konishi operators and describe the kinematics. In Sec.~\ref{sec:ff}, the FF is defined in terms of the renormalised matrix elements. The regularisation prescription is discussed in Sec.~\ref{sec:reg}. The UV operator renormalisation and infrared factorisation in terms of universal quantities are described in Sec.~\ref{ss:op-ren} and \ref{ss:ir}, respectively. The methodology of performing two-loop computation is described in Sec.~\ref{sec:method}. We present our results in Sec.~\ref{sec:res} and discuss about the symmetries which the amplitudes exhibit. We introduce the finite remainder function in Sec.~\ref{sec:finrem}. In the next Sec.~\ref{sec:UT}, we elaborate on the violation of the uniform transcendental property. Finally we make concluding remarks in Sec.~\ref{sec:concl}. The 2-loop results are presented in an ancillary file, \anci, with the \arXiv~submission.

\section{Theoretical Framework}
\label{sec:th}
The Lagrangian describing the dynamics of ${\cal N}=4$ SYM  reads~\cite{Brink:1976bc,Gliozzi:1976qd,
Jones:1977zr,Poggio:1977ma}
\begin{align}
    \mathcal{L}_{{\cal N} = 4} = &-\frac{1}{4}G_{\mu\nu}^a G^{\mu\nu a} - \frac{1}{2\xi}(\partial_{\mu}A^{a\mu})^2 + \partial_{\mu}\bar{\eta}^a 
D^{\mu}\eta_a + \frac{i}{2}\bar{\lambda}^a_m\gamma^{\mu}D_{\mu}\lambda^a_m + \frac{1}{2}(D_{\mu}\phi^a_i)^2 
\nonumber\\
&+ \frac{1}{2}(D_{\mu}\chi^a_i)^2 - \frac{g}{2}f^{abc}\bar{\lambda}^a_m[\alpha^i_{m,n}\phi^b_i +
 \gamma_5\beta^i_{m,n}\chi^b_i]\lambda^c_n - \frac{g^2}{4}\Big[(f^{abc}\phi^b_i\phi^c_j)^2 
\nonumber\\
&+ (f^{abc}\chi^b_i\chi^c_j)^2 + 2 (f^{abc}\phi^b_i\chi^c_j)^2\Big]
\end{align}
where $A,\eta,\lambda,\phi$ and $\chi$ represent the gauge, ghost, Majorana, scalar and pseudo-scalar fields, respectively. All of these transform in the adjoint representation of SU(N) gauge group which is reflected through the presence of the indices $a,b,c$ on the fields. $\xi$ denotes the gauge fixing parameter and $g$ is the Yang-Mills coupling constant. The  indices $m,n \in [1,4]$ represent the four generations of Majorana fermions. The different generations of scalars and pseudo-scalars are captured through $i,j \in [1,n_g]$ with $n_g=3$ in 4-dimensions. $G$ is the gluonic field strength tensor. The fully anti-symmetric structure constants of the SU(N) gauge group are defined through $[T^a,T^b]_{-}\equiv if^{abc} T^c$ and $Tr(T^a T^b)=N \delta^{ab}$ where $T$ are the generators. $\alpha$ and $\beta$ are the anti-symmetric matrices satisfying following algebras:
\begin{align}
    [\alpha^i,\alpha^j]_+=[\beta^i,\beta^j]_+=-2\delta^{ij}, \quad [\alpha^i,\beta^j]_-=0
\end{align}
and
\begin{align}
    tr(\alpha^i)=tr(\beta^j)=tr(\alpha^i\beta^j)=0,\quad tr(\alpha^i\alpha^j)=tr(\beta^i\beta^j)=-4\delta^{ij}\,.
\end{align}

We consider two composite operators, a half-BPS operator~\cite{Bergshoeff:1980is,vanNeerven:1985ja}, ${\cal O}_{\rm hBPS}$, belonging to the stress-energy supermultiplet containing the conserved currents of ${\cal N}=4$ SYM and a non-BPS operator, namely, the Konishi, ${\cal O}_{\rm nBPS}$, which are given by
\begin{align}
\label{eq:op-def}
    &{\cal O}^{\rm hBPS}_{ij}=\phi^a_i\phi^a_j-\frac{1}{3} \delta_{ij}\phi^a_k\phi^a_k\,,\nonumber\\
    &{\cal O}^{\rm nBPS}=\phi^a_i\phi^a_i+\chi^a_i\chi^a_i\,.
\end{align}
More specifically, we are interested in the four-point amplitude of two different off-shell states ($J$) described by these two operators which are produced from two on-shell scalar particles i.e.
\begin{align}
\label{process}
    \phi^a_i(p_1)+\phi^b_j(p_2) \rightarrow J^{\rm hBPS}_{ij}(q_1)+J^{\rm nBPS}(q_2),
\end{align}
where $p_i$ and $q_i$ are the corresponding 4-momentum with $p_i^2= 0$ and $q_i^2= m_i^2$ and the corresponding Madelstam variables are defined as 
\begin{align}
    s\equiv (p_1+p_2)^2\,,~~ t\equiv (p_1-q_1)^2\, ~~\text{and}~~ u\equiv (p_2-q_1)^2,
\end{align}
satisfying $s+t+u=q_1^2+q_2^2=m_1^2+m_2^2$. The physical region of the phase space is bounded by $tu=q_1^2q_2^2$ such that it satisfies
\begin{align}
    s \ge \left( \sqrt{q_1^2}+\sqrt{q_2^2}\right)^2\,, ~~~~ \frac{1}{2} \left(q_1^2+q_2^2-s-\kappa\right) \le t \le \frac{1}{2} \left(q_1^2+q_2^2-s+\kappa\right),
\end{align}
where $\kappa$ is the K\"all\'en function defined as
\begin{align}
    \kappa(s,q_1^2,q_2^2)\equiv \sqrt{s^2+q_1^4+q_2^4-2(s q_1^2+q_1^2 q_2^2+s q_2^2)}\,.
\end{align}
The underlying Lagrangian encapsulating the interaction of these off-shell states to the fields of ${\cal N}=4$ SYM is given by
\begin{align}
{\cal L}_{\rm int} = J^{\rm hBPS}_{ij} {\cal O}^{\rm hBPS}_{ij} + J^{\rm nBPS} {\cal O}^{\rm nBPS}\,.
\end{align}

\section{\texorpdfstring{Four-point Amplitudes for $\phi\phi\rightarrow J^{hBPS}(m_1)J^{nBPS}(m_2)$}{}}
\label{sec:ff}
In order to represent the four-point scattering amplitude, we define the form factors ${\cal F}=1+\sum_{n=1}^{\infty} a^n {\cal F}^{(n)}$ where the component at ${\cal O}(a^n)$ is connected to the matrix elements as
\begin{align}
\label{eq:ff}
    {\cal F}^{(n)} \equiv \frac{\langle {\cal M}^{(0)}|{\cal M}^{(n)} \rangle }{\langle {\cal M}^{(0)}|{\cal M}^{(0)} \rangle}\,.
\end{align}
In the above expression, $|{\cal M}^{(n)}\rangle$ is the $n$-th loop transition matrix element of the production of two off-shell particles having different masses from the on-shell states composed of two colored scalar particles and $a$ is the 't Hooft coupling~\cite{Bern:2005iz} given by
\begin{align}
    a \equiv \frac{g^2 N}{(4\pi)^2} (4 \pi e^{-\gamma_E})^{\epsilon}
\end{align}
where the $\gamma_E\approx 0.5772$ is the Euler constant and $N$ is the Casimir of SU(N) group in adjoint representation. The goal of this article is to compute the  FF at one and two loops i.e. ${\cal F}^{(1)}$ and ${\cal F}^{(2)}$. In the subsequent sections, we describe the methodology of computing these quantities.

\section{Regularisation Prescription}
\label{sec:reg}
The ${\cal N}=4$ SYM is UV finite in 4-dimensions i.e. the $\beta$-function vanishes to all orders in perturbation theory. This ensures the UV finiteness of the on-shell amplitudes and the FFs of the protected operators, like ${\cal O}^{\rm hBPS}$. However, the unprotected operators like ${\cal O}^{\rm nBPS}$, do develop UV divergences arising from short distance effects. In addition to that, this theory is not free from soft and collinear divergences (IR) due to the presence of massless fields. The on-shell amplitudes, FF of protected as well as unprotected operators give rise to these IR divergences. In order to identify these divergences, we need to regularise the theory. 

Among many other regularisation schemes, we employ the ${\overline{\rm DR}}$ scheme for our computation. This is very much similar to the dimensional regularisation by 't Hooft and Veltman~\cite{tHooft:1972tcz} and preserves SUSY. In order to do that the number of fermionic and bosonic degrees of freedom (DOF) should be maintained equal throughout the calculation. This is achieved by changing the number of generations of the scalar and pseudo-scalar from $n_g=3$ to $n_{g,\epsilon}=3+\epsilon$ in $d=(4-2\epsilon)\equiv d_{\epsilon}$ space-time dimensions. Along with the $(2-2\epsilon)$ DOF of the gauge fields, the total number of bosonic DOF becomes $8$ which is equal to the fermionic ones, same as that of $4$-dimensions and consequently, preserves SUSY. In this scheme, all the traces of the $\alpha$, $\beta$ and Dirac matrices are performed in $d_{\epsilon}$ dimensions. In addition to the usual Lie algebra obeyed by the six anti-symmetric matrices $\alpha$ and $\beta$, they fulfil
\begin{align}
    \alpha^i\alpha^i=\beta^i\beta^i=(-3-\epsilon)\mathbb{I},~~ \alpha^i\alpha^j\alpha^i=\alpha^j(1+\epsilon)\mathbb{I},~~
    \beta^i\beta^j\beta^i=\beta^j(1+\epsilon)\mathbb{I}
\end{align}
in $d_{\epsilon}$ dimensions. Unlike other available schemes, like FDH scheme, the $\overline{DR}$ scheme is universally applicable to the amplitudes for the protected as well as unprotected operators. 

To maintain the traceless property of the ${\cal O}^{\rm hBPS}$, defined in Eq.~\ref{eq:op-def}, in $d_{\epsilon}$ dimensions, it has to be modified as
\begin{align}
    {\cal O}^{\rm hBPS}_{ij}=\phi^a_i\phi^a_j-\frac{1}{n_{g,\epsilon}} \delta_{ij}\phi^a_k\phi^a_k\,.
\end{align}

\subsection{UV Divergences and Operator Renormalisation}
\label{ss:op-ren}
As mentioned in the previous Sec.~\ref{sec:reg}, the amplitudes of the unprotected composite operators like ${\cal O}^{\text{nBPS}}$ are not UV finite which can be seen by the presence of non-zero anomalous dimensions, $\gamma^{\cal K}$,~\cite{Eden:2000mv,ANSELMI1997221,Bianchi:2000hn,Kotikov:2004er,Eden:2004ua,Bajnok:2008bm,Bajnok:2009vm,Fiamberti:2007rj,Fiamberti:2008sh,Velizhanin:2008jd,Eden:2012fe} in the FF. So, any amplitude involving unprotected operator needs to go through UV renormalisation which is performed by multiplying an overall operator renormalisation constant, $Z^{\cal K}(a(\mu^2),\epsilon)$, defined through
\begin{align}
\label{eq:rg}
    \frac{d}{d\ln \mu^2} \ln Z^{\cal K}(a(\mu^2),\epsilon)=\gamma^{\cal K}=\sum_{i=1}^{\infty} a^i(\mu^2) \gamma^{\cal K}_i\,.
\end{align}
This is totally different than the coupling constant or wave-function renormalisation and is solely arising due to the nature of the composite operator. The scale $\mu$ appeared in the above renormalisation group (RG) equation is introduced through
\begin{align}
    {\hat a}=a(\mu^2)  \left(\frac{\mu^2}{\mu_0^2} \right)^{\epsilon},
\end{align}
where, ${\hat a}$ is the coupling constant appearing in the regularised Lagrangian in $d_{\epsilon}$ dimensions and $\mu_0$ is the scale introduced to make ${\hat a}$ dimensionless. Because of the vanishing $\beta$-functions in ${\cal N}=4$ SYM, the coupling constant $a$ satisfies very simple RG equation
\begin{align}
    \frac{d}{d \ln \mu^2}\ln a(\mu^2)=-\epsilon\,.
\end{align}
Employing this, the RG equation for $Z^{\cal K}$, Eq.~\ref{eq:rg}, can be solved exactly to all orders which is obtained as
\begin{align}
    Z^{\cal K}(a(\mu^2),\epsilon)=\exp \left(- \sum_{n=1}^{\infty} a^n(\mu^2) \frac{\gamma^{\cal K}_n}{n\epsilon} \right)
\end{align}
with $\gamma^{\cal K}_1=-6$ and $\gamma^{\cal K}_2=24$. The $Z^{\cal K}$ up to two loops reads
\begin{align}
    Z^{\cal K}(a(\mu^2),\epsilon)=1+a(\mu^2)\frac{6}{\epsilon}+a^2(\mu^2) \left( \frac{18}{\epsilon^2}-\frac{12}{\epsilon}\right)\,.
\end{align}
The UV renormalised matrix element can be written in terms of the bare quantities as
\begin{align}
\label{eq:un}
    |{\cal M}\rangle &= Z^{\cal K}  \sum_{n=0}^{\infty} \left( {\hat a} \mu_0^{2\epsilon} \right)^n |{\hat {\cal M}}^{(n)}\rangle
\end{align}
where, $|{\hat {\cal M}}^{(n)}\rangle$ is the $n$-th loop bare amplitude. On the other hand, we can equally express $|{\cal M}\rangle$ in powers of renormalised quantities $|{\cal M}^{(n)}\rangle$ through
\begin{align}
\label{eq:ren}
    |{\cal M}\rangle =  \sum_{n=0}^{\infty} \left( a(\mu^2) \mu^{2\epsilon} \right)^n |{\cal M}^{(n)}\rangle\,.
\end{align}
By comparing the Eq.~\ref{eq:ren} with \ref{eq:un}, we get the UV renormalised matrix elements in terms of bare ones as
\begin{align}
    |{\cal M}^{(0)}\rangle &= |{\hat {\cal M}}^{(0)}\rangle\,,\nonumber\\
    |{\cal M}^{(1)}\rangle &= \mu^{4\epsilon}|{\hat {\cal M}}^{(1)}\rangle -\mu^{2\epsilon} \frac{\gamma_1^{\mathcal{K}}}{\epsilon} |{\hat {\cal M}}^{(0)}\rangle \,,\nonumber\\
    |{\cal M}^{(2)}\rangle &= \mu^{6\epsilon}|{\hat {\cal M}}^{(2)}\rangle -\mu^{4\epsilon} \frac{\gamma_1^{\mathcal{K}}}{\epsilon} |{\hat {\cal M}}^{(1)}\rangle + 
    \mu^{2\epsilon} \left( \frac{\left(\gamma^{\mathcal{K}}_1\right)^2}{2\epsilon^2}-\frac{\gamma_2^{\mathcal{K}}}{2\epsilon}\right) |{\hat {\cal M}}^{(0)}\rangle\,.
\end{align}
Substituting the above relations in Eq.~\ref{eq:ff}, we obtain the UV renormalised form factors at 1- and 2-loop level which are presented in this article.

\subsection{Universality of IR Divergences}
\label{ss:ir}
Beyond leading order, the resulting UV renormalised FF contains IR divergences arising from the soft and collinear configurations which appear as poles in the dimensional regularisation parameter $\epsilon$. Due to the universal nature of these singularities, these are process independent and depend only on the nature of the external particles containing SU(N) color index. In a seminal paper~\cite{Catani:1998bh}, Catani predicted these poles for $n$-point two loops scattering amplitudes which are related to universal anomalous dimensions. Later, a formal derivation was presented in~\cite{Sterman:2002qn} exploiting the factorisation and resummation properties of QCD amplitudes, and were subsequently generalised to all loop order in~\cite{Becher:2009cu,Gardi:2009qi}. Following~\cite{Catani:1998bh}, we get
\begin{align}
\label{eq:mat-cat}
    |{\cal M}^{(1)}\rangle &= 2 {\mathbf{I}}^{(1)}(\epsilon) |{\cal M}^{(0)}\rangle + |{\cal M}^{(1)}_{\rm fin}\rangle\,,\nonumber\\
    |{\cal M}^{(2)}\rangle &= 4 {\mathbf{I}}^{(2)}(\epsilon) |{\cal M}^{(0)}\rangle + 2 {\mathbf{I}}^{(1)}(\epsilon) |{\cal M}^{(1)}\rangle + |{\cal M}^{(2)}_{\rm fin}\rangle\,.
\end{align}
In the context of ${\cal N}=4$ SYM where all the fields are in the adjoint representation and $\beta$-function vanishes to all orders, the IR subtraction operators are obtained by keeping only the highest transcendental terms which turn out to be
\begin{align}
\label{eq:sub}
    &{\mathbf{I}}^{(1)}(\epsilon)=-\frac{e^{\epsilon\gamma_E}}{\Gamma(1-\epsilon)} \left( -\frac{\mu^2}{s} \right)^{\epsilon} \left(\frac{1}{\epsilon^2}\right)\,,\nonumber\\
    &{\mathbf{I}}^{(2)}(\epsilon)=-\frac{1}{2} \left(\mathbf{I}^{(1)}(\epsilon)\right)^2 - e^{-\epsilon\gamma_E}\frac{ \Gamma(1-2\epsilon)}{\Gamma(1-\epsilon)} \zeta_2~ \mathbf{I^{(1)}}(2\epsilon) + \frac{\mathbf{H}^{(2)}(\epsilon)}{\epsilon}
\end{align}
with~\cite{Becher:2009cu}
\begin{align}
    \mathbf{H}^{(2)}(\epsilon)=\frac{e^{\epsilon\gamma_E}}{\Gamma(1-\epsilon)} \l( -\frac{\mu^2}{s}\r)^{2 \epsilon}\frac{1}{4}\zeta_3.
\end{align}
The finite i.e. ${\cal O}(\epsilon^{\alpha}), ~\alpha\geq 0$ terms in the subtraction operators are arbitrary and these define the scheme in which the finite part of the amplitude, $| {\cal M}^{(n)}_{\rm fin}\rangle$, is computed. Translating the infrared structures of the matrix elements to the UV renormalised form factors in Eq.~\ref{eq:ff}, we can write the IR finite ${\cal F}_{\rm fin}^{(n)}$. For the convenice of the readers, we write these below:
\begin{align}
\label{eq:fffin}
    &{\cal F}^{(1)}(\epsilon)=2 {\mathbf{I}}^{(1)}(\epsilon) + {\cal F}^{(1)}_{\rm fin}\,,\nonumber\\
    &{\cal F}^{(2)}(\epsilon)=4 {\mathbf{I}}^{(2)}(\epsilon) + 2 {\mathbf{I}}^{(1)}(\epsilon) {\cal F}^{(1)}(\epsilon) + {\cal F}_{\rm fin}^{(2)}\,.
\end{align}
The goal of this article is to compute the quantities ${\cal F}^{(1)}_{\rm fin}$ and ${\cal F}^{(2)}_{\rm fin}$. In the next section, we describe the methodology for the computation of the four-point amplitudes.

\section{Calculation of the Amplitudes}
\label{sec:method}
In contrast to the most popular method of unitarity for computing the scattering amplitudes in the context of ${\cal N}=4$ SYM, we employ the Feynman diagrammatic approach that carries advantages in light of the regularisation scheme. The Feynman diagrams for the process under consideration are generated using \qgraf~\cite{Nogueira:1991ex}. Special care is taken in order to incorporate the Majorana fermions which are its own anti-particle and consequently, destroy the flow of fermionic current in the \qgraf~ output. This is rectified by an in-house code based on \Python. There are 2, 18 and 489 Feynman diagrams at tree, one and two loop, respectively. The \qgraf~ output is passed through several in-house codes written in symbolic manipulating program \form~\cite{Vermaseren:2000nd} in order to apply the Feynman rules, perform Dirac, Lorentz and SU(N) color algebras. We employ the Feynman-'t Hooft gauge ($\xi=1$) for the internal gluons. After evaluation of traces of Dirac's $\gamma$-matrices and contraction of Lorentz indices every Feynman diagram is expressed as linear combination of a larger number of scalar Feynman integrals which belong to the family of the massless four-point functions with two off-shell legs of different virtualities. Using the liberty of transforming the loop momenta, all the scalar integrals are categorised into 3 and 6 different integral families at 1- and 2-loop, respectively, with the help of \reduze~\cite{vonManteuffel:2012np,Studerus:2009ye}. These scalar integrals are reduced to a smaller set of master integrals (MI) employing the integration-by-parts (IBP)~\cite{Tkachov:1981wb,Chetyrkin:1981qh} and Lorentz invariant~\cite{Gehrmann:1999as} identities. Being a process involving two massive external legs of different masses, the reduction is resonably complicated. In order to achieve that, we use \litered~\cite{Lee:2008tj,Lee:2012cn} along with \mint~\cite{Lee:2013hzt} at 1-loop and \kira~\cite{Maierhoefer:2017hyi,Maierhofer:2018gpa} at 2-loop. Moreover, we cross-check the 2-loop reductions with the c++ version of \fire~\cite{Smirnov:2014hma}. 

Upon performing the IBP reductions, there are 134 MIs at 2-loop level. Among these, 49 are related to each other by crossing and 2-integrals turn out to be same which are related to each other by simple transformation of the loop momentum. At the end, we have 84 independent master integrals. These are matched with the set of MIs already present in the literature. The integrals for the non-equal masses were first derived in~\cite{Henn:2014lfa,Caola:2014lpa,Papadopoulos:2014hla}. A subset of these MIs were computed in~\cite{Chavez:2012kn,Anastasiou:2014nha}. An independent derivation of these MIs was performed in~\cite{Gehrmann:2015ora} where the solutions are optimised for the numerical evaluation. For our current calculation, we use the optimised solutions of the MIs presented in~\cite{Gehrmann:2015ora} which are available in \hepforge~\cite{hepforge} in computer readable format. To get the optimised solutions of the MIs~\cite{Gehrmann:2015ora}, a convenient choice of variables $x,y,z$ and $m^2$ is  made:
\begin{align}
\label{eq:varch}
    s=m^2(1+x)(1+xy)\,,~~~ t=-m^2xz\,,~~~ q_1^2=m^2\,,~~~ q_2^2=m^2x^2y,
\end{align}
which rationalise the root of $\kappa$. The physical region is constrained through
\begin{align}
    x>0\,,~~~ 0<y<z<1\,,~~~ m^2>0\,.
\end{align}
The symbol alphabet of the MIs involve 19 letters
\begin{align}
\label{eq:let}
    l_i \in \{ &x, 1 + x, y, 1 - y, z, 1 - z, -y + z, 1 + y - z, 1 + xy, 1 + xz, xy + z,\nonumber\\ &1 + y + xy - z, 1 + x + xy - xz, 1 + y + 2xy - z + x^2yz,\nonumber\\ &2xy+x^2y+x^2y^2 +z-x^2yz,1+x+y+xy+xy^2 -z-xz-xyz,
   \nonumber\\ &1+y+xy+y^2 +xy^2 -z-yz-xyz, -xy + z + xz + xyz, 
    \nonumber\\ &-y + z + yz + xyz\}\,.
\end{align}
The results are presented as Laurent series expansion in $\epsilon$ to weight 4 which enable us to obtain the 2-loop results to ${\cal O}(\epsilon^0)$. In the next section~\ref{sec:res}, we discuss and present the results.

\section{Results of the One \& two loops Amplitudes}
\label{sec:res}
We calculate the 1- and 2-loop form factors, Eq.~\ref{eq:ff}, to ${\cal O}(\epsilon^2)$ and ${\cal O}(\epsilon^0)$ where the results contain maximum weight 4 terms. The results are expressed in terms of logarithms $\log$, classical polylogarithms ${\rm Li}_n~(n=2,3,4)$ and ${\rm Li}_{2,2}$ where the latter one is defined in \cite{Gehrmann:2015ora}. The square of the Born amplitude is obtained as
\begin{align}
\label{eq:bornsq}
    \langle{\cal M}^{(0)}|{\cal M}^{(0)}\rangle = (N^2-1) \frac{20(1+y)^2}{m^4 x^2 z^2 (1+y-z)^2}\,.
\end{align}
The above expression is symmetric under the exchange of $z \leftrightarrow (1+y-z)$ which translates to $t \leftrightarrow u$ or equivalently $p_1 \leftrightarrow p_2$. This is in accordance with the expectation. In the next subsection~\ref{ss:symm}, we elaborate on this and present the results for 1-loop and 2-loop partially. Complete results can be found from the ancillary file supplied with the \arXiv~ submission in Mathematica format.

\subsection{Permutation Symmetry}
\label{ss:symm}
\begin{figure}[ht]
    \centering
    \includegraphics[scale=0.265]{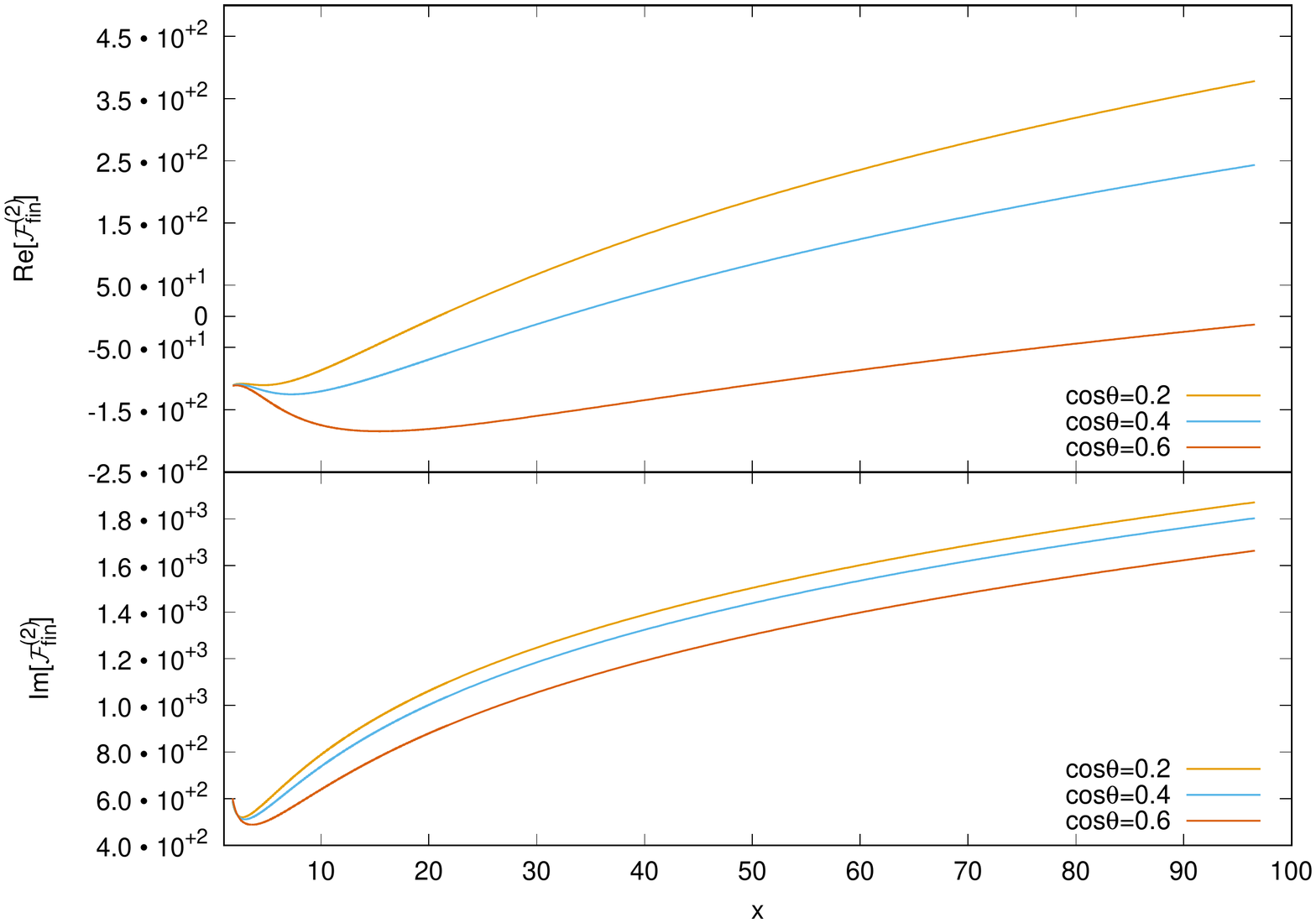}
    \includegraphics[scale=0.265]{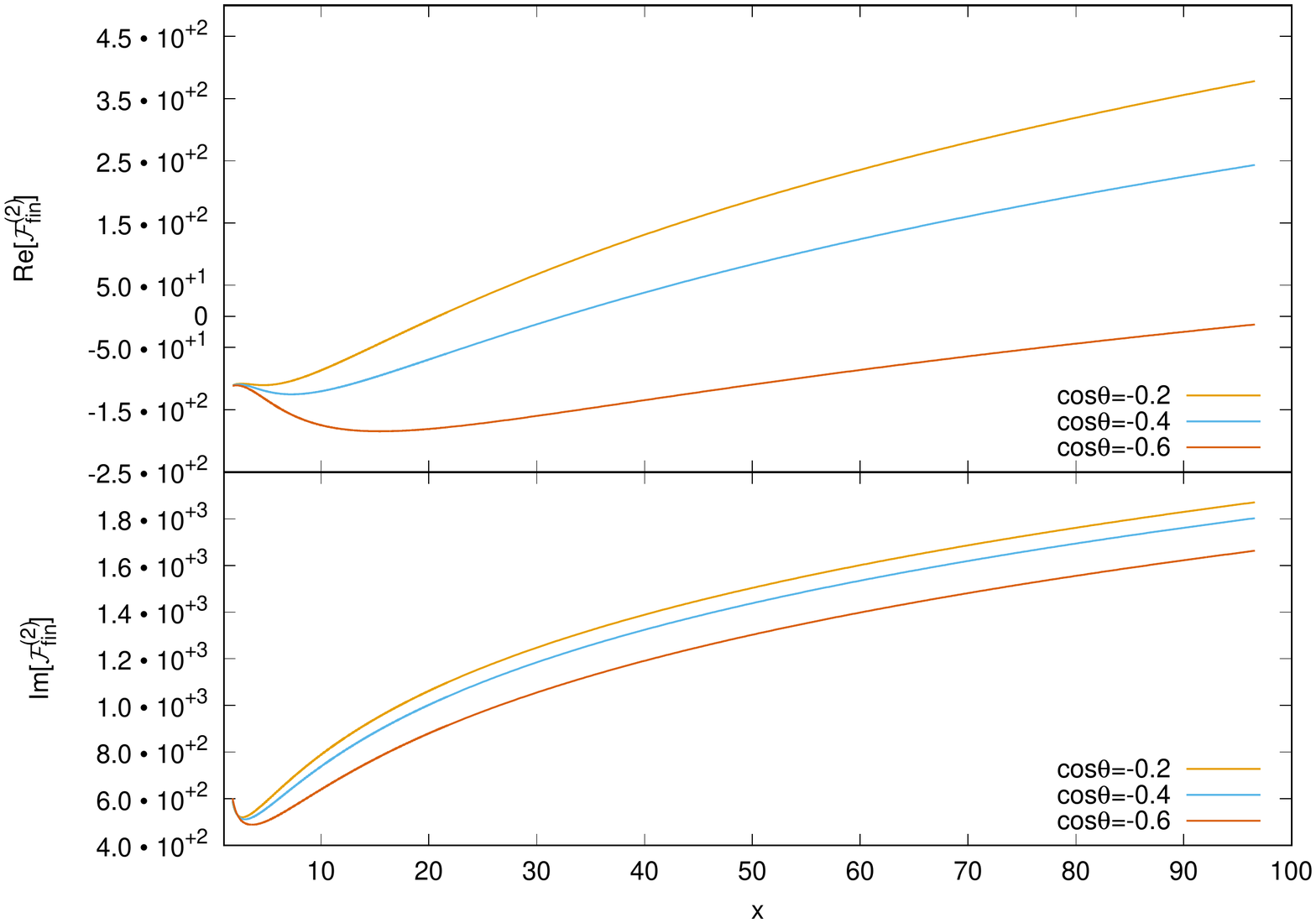}
    \caption{Behaviour of the real and imaginary parts of ${\cal{F}}_{\rm fin}^{(2)}$ as a function of variable $x$ for different values of $\cos{\theta}$.}
    \label{fig:tusymm}
\end{figure}

The process under consideration is symmetric under the exchange of initial state particles that is under $p_1 \leftrightarrow p_2$ which translates to the Mandelstam variables as
\begin{align}
    &t \leftrightarrow u \Rightarrow z \leftrightarrow 1+y-z\,.
\end{align}
This symmetry should be reflected in the amplitude. Since the results are expressed in terms of functions containing variables defined through Eq.~\ref{eq:varch}, it is not straightforward to check the symmetry analytically. In order to check this, we go to a new set of variables, $\T$ and $\U$ introduced through   
\begin{align}
    &\T \equiv z\,,~ \U \equiv 1+y-z \Rightarrow z=\T\,,~y=\U+\T-1\,.
\end{align}
The permutation symmetry implies the amplitude should remain invariant under the exchange of $\T \leftrightarrow \U$.
In terms of these new variables, the born amplitude Eq.~\ref{eq:bornsq} becomes
\begin{align}
\label{eq:bornsq1}
    \langle{\cal M}^{(0)}|{\cal M}^{(0)}\rangle = (N^2-1)\frac{20 (\T+\U)^2}{m^4 x^2 \T^2 \U^2},
\end{align}
where the $\T \leftrightarrow \U$ symmetry is explicit. Note that the leading order amplitude does depend on $\epsilon$. In above Eq.~\ref{eq:bornsq1}, only the ${\cal O}(\epsilon^0)$ term is presented. To check this behaviour for 1- and 2-loop amplitudes, we express the letters in Eq.~\ref{eq:let} in terms of these variables which turn out to be
\begin{align}
    l_i \in \{&x, 1 + x, \T + \U-1, 2 - \T - \U, \T, 1 - \T, 1 - \U, \U, 1 + x (\T + \U-1),
 1 + \T x,\nonumber\\ &\T +  x (\T + \U-1), \U +  x (\T + \U-1), 1 + \U x,\nonumber\\&
 \U + 2  x (\T + \U-1) +  x^2 \T (\T + \U-1), \nonumber\\&\T + 2 x (\T + \U-1) +
   x^2 \U (\T + \U-1), \U + x - \T x + x \U (\T + \U-1),\nonumber\\&
 1 + \U (\U-1)  (1 + x) + \T (\U + \U x-1), x +\T+ x (\T-1) (\T + \U),\nonumber\\&
 1 - \U + \T (\T + \U-1) (1 + x)\}\,.
\end{align}
We see under the exchange of $\T \leftrightarrow \U$, some letters transform among themselves:
\begin{align}
    &l_5 \leftrightarrow l_8,~ l_6 \leftrightarrow l_7,~ l_{10} \leftrightarrow l_{13},~ l_{11}\leftrightarrow l_{12},~ l_{14} \leftrightarrow l_{15},~ l_{16} \leftrightarrow l_{18},~ l_{17} \leftrightarrow l_{19},
\end{align}
whereas the remaining ones are unaffected. Using these, we see the 1-loop form factor indeed remains invariant under the exchange of $p_1\leftrightarrow p_2$. By making this behaviour explicit, we present the ${\cal O}(\epsilon^0)$ term of the 1-loop finite part (related to the hard function in QCD), Eq.~\ref{eq:fffin}, below
\begin{align}
\label{eq:ff1l}
    {\cal F}^{(1)}_{\rm fin} &=\bigg\{-7-3i\pi + 6\log l_1+3\log l_3-\frac{1}{3}\pi^2 - 2i\pi \log\left(l_2l_9\right) + \log l_3\log(l_2l_9)
    \nonumber\\
    &+ \frac{2u}{t+u}\bigg[\log^2l_{10} 
    +\log^2l_{11}-\log^2l_5-2\log l_{11}\log\left(l_1l_3 \right)+2\log l_5\log\left(\frac{l_1l_3}{l_2l_9}\right)
    \nonumber\\
    &+2\text{Li}_2\left(\frac{1}{l_{10}}\right)
    +2\text{Li}_2\left(\frac{l_1l_3}{l_{11}}\right)
    + 2i\pi \log\left(\frac{l_{10}l_{11}}{l_5}\right)  \bigg] \bigg\} + \bigg\{ t \leftrightarrow u \bigg\}\,.
\end{align}
The $\{t \leftrightarrow u\}$ represents the terms obtained by performing the interchange of $t$ and $u$ in the first part of the expression in Eq.~\ref{eq:ff1l}. 
The ${\cal O}(\epsilon^0)$ term of the 2-loop finite part of the FF, Eq.~\ref{eq:fffin}, can be written as
\begin{align}
\label{eq:ff2lexpand}
    {\cal F}^{(2)}_{\rm fin} = \sum_{k=0}^4 {\cal F}^{(2),\tau(k)}_{\rm fin}
\end{align}
where $\tau(k)$ represents the transcendentality of degree $k$, defined in the introduction, Sec.~\ref{sec:intro} and ${\cal F}^{(2),\tau(k)}$ represents the terms having transcendentality $k$. Due to relatively large size of the terms, we present only the lower transcendental weight terms at 2-loop which read
\begin{align}
\label{eq:ff2l}
    {\cal F}^{(2),\tau(3)}_{\rm fin} &= \frac{u}{t+u}\Bigg[4i\pi^3 -4\pi^2\left\{\log \left(l_1^2l_3\right)+6\log(l_2l_9)-6\log\left(\frac{l_{10}l_{11}}{l_5}\right) \right\}
    \nonumber\\&
    -12i\pi\bigg\{ \log^2l_{10}+\log^2l_{11}-\log^2l_5+4\log l_1\log\left(\frac{l_2l_9}{l_{10}}\right)+6\log l_1\log\left(\frac{ l_5}{l_{11}}\right)
    \nonumber\\&
    +4\log l_3\log\left(\frac{l_5}{l_{11}}\right)+3\log l_3 \log\left(l_2l_9\right)-2\log \left(l_2l_9\right)\log l_5 - 2\log l_3 \log l_{10}
    \nonumber\\&
    +2\text{Li}_2\left(\frac{1}{l_{10}}\right)+2\text{Li}_2\left(\frac{l_1l_3}{l_{11}}\right)\bigg\}
    +12\log^2l_3\log\left(l_2l_9\right)+48\log^2l_1\log\left(\frac{l_5}{l_{11}}\right)
    \nonumber\\&
    +24\log^2l_3\log\left(\frac{l_5}{l_{11}}\right)+12\log\left(l_1^2l_3\right)\left\{\log^2l_{10}+\log^2l_{11}-\log^2l_5 \right\} 
    \nonumber\\&
    + 72\log l_1 \log l_3 \log\left(\frac{l_5}{l_{11}} \right) + 24\log l_3 \log\left(l_2l_9\right)\log\left(\frac{l_1}{l_5} \right)-48\log l_1\log l_5 \log\left(l_2l_9\right)
    \nonumber\\&
    +24\log\left(l_1^2l_3\right)\left\{\text{Li}_2\left(\frac{1}{l_{10}} \right) +\text{Li}_2\left(\frac{l_1l_3}{l_{11}} \right) \right\}\Bigg]+ \bigg\{ t \leftrightarrow u\bigg\}\,,
\nonumber\\
     {\cal F}^{(2),\tau(2)}_{\rm fin} &= \Bigg[-\frac{5n_1}{3d_1}\pi^2+i\pi\bigg\{28\log\left( l_2l_9\right)+\frac{u}{t+u}\left(56\log l_5-32\log l_{11}\right)-48\log l_1 -\frac{4u n_5}{d_2}
    \nonumber\\&
    \times\log l_{10}-\frac{6n_4}{d_1}\log l_3 + \frac{24n_3}{d_1}\log l_6\bigg\} +42\log^2l_1 + \frac{u}{t+u}\left(28\log^2l_5-16\log^2l_{11} \right)
    \nonumber\\&
    -\frac{2u n_5}{d_2}\log^2l_{10}+\frac{3n_2}{d_1}\log^2l_3 + \frac{6n_3}{d_1}\log^2l_6+\frac{32u}{t+u}\log l_{11}\log(l_1l_3)+48\log l_1\log l_{3}
    \nonumber\\&-14\log l_3\log(l_2l_9)-\frac{56u}{t+u}\log l_5\log\left( \frac{l_1l_3}{l_2l_9}\right)-\frac{12n_3}{d_1}\log l_6\log\left(\frac{l_3}{l_5} \right)+\frac{12n_3}{d_1}\text{Li}_2( l_5)
    \nonumber\\&
    -\frac{32u}{t+u}\text{Li}_2\left(\frac{l_1l_3}{l_{11}} \right)-\frac{4u n_5}{d_2}\text{Li}_2\left(\frac{1}{l_{10}} \right) + \frac{12n_3}{d_1}\text{Li}_2\left(-\frac{l_3}{l_6} \right)\Bigg] + \bigg\{t\leftrightarrow u \bigg\}\,,\nonumber\\
     {\cal F}^{(2),\tau(1)}_{\rm fin} &=108 i \pi-216 \log (l_1)-108 \log (l_3)\,,\nonumber\\
     {\cal F}^{(2),\tau(0)}_{\rm fin} &= 204
\end{align}
where the $\{n_i\}$ and $\{d_i\}$ are given by
\begin{align}
\label{eq:nd}
    n_1&=11 m^4 (t + u) + 5 s^2 (t + u) + 11 t u (t + u) + 
   s (5 t^2 + 22 t u + 5 u^2) \nonumber\\&- m^2 (t + u) (16 s + 11 (t + u))\,,\nonumber\\
   n_2&= 6 m^4 (t + u) + 5 s^2 (t + u) + 6 t u (t + u) + 
   s (5 t^2 + 12 t u + 5 u^2) \nonumber\\&- m^2 (t + u) (11 s + 6 (t + u))\,,\nonumber\\
   n_3&=m^4 (t+u)-m^2 (t+u) (s+t+u)+t u (2 s+t+u)\,,\nonumber\\
   n_4&=7 m^4 (t + u) + 5 s^2 (t + u) + 7 t u (t + u) + 
  s (5 t^2 + 14 t u + 5 u^2) \nonumber\\ &- m^2 (t + u) (12 s + 7 (t + u))\,,\nonumber\\
   n_5&=17 m^2 - 14 s - 17 t\,,\nonumber\\
   d_1&=(m^2 - s - t) (m^2 - s - u) (t + u)\,,\nonumber\\
   d_2&=(m^2 - s - t) (t + u)\,.
\end{align}
The remaining, namely, the highest transcendental term ${\cal F}^{(2),\tau(4)}_{\rm fin}$ can be found in the ancillary file supplied with the \arXiv~submission. The renormalisation scale $\mu$ is set equal to $m$ throughout the article which can be restored by using renormalisation group evolution. The 1-loop form factor is calculated to ${\cal O}(\epsilon^2)$ and 2-loop to ${\cal O}(\epsilon^0)$ or to weight 4 terms. The $t \leftrightarrow u$ symmetry is checked for all these terms. It is checked analytically for the 1-loop terms to ${\cal O}(\epsilon)$ and for the lower transcendental terms of the 2-loop to ${\cal O}(\epsilon^0)$. On the other hand, for the ${\cal O}(\epsilon^2)$ term at 1-loop and the highest transcendental part of the 2-loop form factor, numerical checks are performed which is necessitated by the long expression. In order to demonstrate the $t\leftrightarrow u$ symmetry at 2-loop, we plot in Fig.~\ref{fig:tusymm}, the real and imaginary parts of ${\cal F}_{\rm fin}^{(2)}$ as a function of $x$ for different choices of $\cos{\theta}$, where $\theta$ is the angle between one of the composite operators under consideration and one of the initial scalars in their center of mass frame. Without loss of generality, we set $\sqrt{q_1^2} = 100$ GeV and $\sqrt{q_2^2} = 150$ GeV for numerical purpose. The behaviour of the amplitude is clearly seen to be invariant under $\cos{\theta}\rightarrow -\cos{\theta}$, which reflects the the $t \leftrightarrow u$ symmetry in the final 2-loop expression.
In addition to exhibiting the universal infrared structure, this symmetry serves as a very strong check on our calculation, in particular on the finite parts of the FF.

\section{Finite Remainders}
\label{sec:finrem}
The iterative structure of form factors in terms of Catani's IR subtraction operators up to two loops level is a result of factorisation at long distances. The observed universality of these IR poles lead to exponentiation of these divergences whose coefficients are controlled by universal anomalous dimensions. In the context of the BDS ansatz~\cite{Bern:2005iz}, both the IR divergences and finite parts can be exponentiated and the exponents are expressed in terms of their one-loop counter parts multiplied by a function that depends on the anomalous dimensions such as cusp and collinear ones.  For maximally helicity violating (MHV) amplitudes, this iterative structure of the finite terms breaks down starting from two-loop six-point~\cite{Bartels:2008ce,Bern:2008ap} and the deviation is captured through a quantity called finite remainder function.

In our case of study, the IR divergences exponentiate, thanks to their iterative structure but not the finite terms. Hence the corresponding finite remainder functions are always non-zero. For the two-point half-BPS operator, the authors of~\cite{Brandhuber:2010ad} have obtained the finite reminder function up to two loops. Later it was extended to cases with more than two external states~\cite{Brandhuber:2012vm,Banerjee:2016kri}. Following BDS conjecture, the finite reminder function at two loops is defined as
\begin{align}
\label{eq:fin-rem}
\mathcal{R}^{(2)} = \mathcal{F}^{(2)}(\epsilon)-\frac{1}{2}\left(\mathcal{F}^{(1)}(\epsilon)\right)^2-f^{(2)}(\epsilon)\mathcal{F}^{(1)}(2\epsilon)-C^{(2)},
\end{align}
where $f^{(2)}(\epsilon)=-2\zeta_2-2\epsilon\zeta_3-\frac{4}{5}\epsilon^2\zeta_2^2$ and $C^{(2)}=\frac{8}{5}\zeta_2^2$. Note that $f^{(2)}(\epsilon)$ and $C^{(2)}$ are independent of operators as well as external states which is a consequence of the universality of IR divergences. In this article, we compute the finite remainder function for the amplitude under consideration at two loops. The ${\cal O}(\epsilon^0)$ term of the ${\mathcal{R}}^{(2)}$, Eq.~\ref{eq:fin-rem}, can be written as
\begin{align}
\label{eq:rem2lexpand}
    \mathcal{R}^{(2)} = \sum_{k=0}^4 \mathcal{R}^{(2),\tau(k)}
\end{align}
where $\tau(k)$ represents the transcendentality of degree $k$, defined in the introduction, Sec.~\ref{sec:intro} and $\mathcal{R}^{(2),\tau(k)}$ represents the terms having transcendentality $k$. Due to the large size of the terms, we present only the lower transcendental weight terms. The full result can be found from the ancillary file supplied with the \arXiv~ submission. The lower transcendental terms read as
\begin{align}
    {\cal R}^{(2),\tau(3)} &=
    -2 i \pi^3 + \pi^2 \left(4 \log(l_1) + 2 \log(l_3)\right)\,,\nonumber\\
    {\cal R}^{(2),\tau(2)} &= 
   \bigg\{ -\frac{2\pi^2}{3d_1} \bigg[ 49 {m}^4 t
    -{m^2} t (68 s+49 (t+u))
    +19 s^2 t+s (19 t^2
    +49 t u)+49 t^2 u \bigg] \nonumber\\&
    + \pi \bigg[ -12 i \log(l_1)
    - \frac{24 i t}{d_1} \log(l_3) ({m^2}
    -s-t) (2 {m^2}-s
    -2 u)
    + \frac{24in_3}{d_1} \log(l_6) \nonumber\\&
    -\frac{12i}{d_2} \log(l_{10}) (m^2 - t) u
    +\frac{24iu}{t+u} \log(l_{11})
    \bigg]
    + 6\log^2(l_1) \nonumber\\&
    + \frac{6}{d_1} \log^2(l_3) \bigg( 3 m^4 t + 2 s^2 t 
    + 3 t ^2 u 
    + s (2 t^2 + 3 t u) - 
 m^2 t (5 s + 3 (t + u))\bigg) \nonumber\\&
 +\frac{6n_3}{d_1} \log^2(l_6) 
 - \frac{6}{d_2} \log^2(l_{10}) (m^2 - t) u
 + \frac{12u}{t+u} \log^2(l_{11})
 - \frac{24u}{t+u} \log(l_1l_3) \log(l_{11}) \nonumber\\&
 + 12 \log(l_{1}) \log(l_3)
 -\frac{12n_3}{d_1} \log(l_3) \log(l_6) 
 +\frac{12n_3}{d_1} \log(l_5) \log(l_6) +\frac{24u}{t+u} \text{Li}_2\left(\frac{{l_1} {l_3}}{{l_{11}}}\right)\nonumber\\&
 + \frac{12}{d_2} \text{Li}_2\left(\frac{1}{{l_{10}}}\right) (t-m^2) u
 +\frac{12n_3}{d_1} \text{Li}_2({l_5})
 + \frac{12n_3}{d_1}\text{Li}_2\left(-\frac{{l_3}}{{l_6}}\right) \bigg\} + \bigg\{ t \leftrightarrow u \bigg\}\,,\nonumber\\
 {\cal R}^{(2),\tau(1)} &=
     24 i \pi-48 \log ({l_1})-24 \log ({l_3})\,,\nonumber\\
     {\cal R}^{(2),\tau(0)} &=
     106\,,
\end{align}
where the $n_i$ and $d_i$ are defined in Eq.~\ref{eq:nd}. Finite remainders are also checked to obey the $t \leftrightarrow u$ symmetry. Owing to a large number of cancellation among all the terms, the ${\tau(3)}$ terms of the finite reminder turns out to be surprisingly small compared to other terms. In the next section, we discuss the behaviour of highest transcendentality terms.

\section{Behaviour of Leading Transcendental Terms}
\label{sec:UT}
It is an observed~\cite{Kotikov:2002ab,Kotikov:2004er,Bern:2006ew,Drummond:2007cf,Naculich:2008ys,Bork:2010wf,Gehrmann:2011xn,Brandhuber:2012vm,Eden:2012rr,Drummond:2013nda,Basso:2015eqa,Goncalves:2016vir,Banerjee:2018yrn}, albeit unproven fact that certain kinds of scattering amplitudes in ${\cal N}=4$ SYM exhibit uniform transcendentality (UT) i.e. those can be expressed in terms of polylogarithmic functions of uniform degree 2L with constant coefficients, where L denotes the loop order. However, this property is no longer true for Sudakov~\cite{Nandan:2014oga,Ahmed:2016vgl} or three point~\cite{Banerjee:2016kri} form factor of non-protected operators. Naturally, it is expected that the four point amplitude involving a half-BPS and Konishi operator would involve highest (2L) as well as lower ($<$2L) transcendental terms. Our explicit calculation to two loops shows, Eq.~\ref{eq:ff1l} and~\ref{eq:ff2lexpand}, it is indeed true. 

Moreover, looking at the one loop finite FF in Eq.~\ref{eq:ff1l} reveals that the coefficients of the highest transcendental terms are indeed simple numerical constant apart from an overall kinematic factor. This is a reflection of the very simple structure of the amplitude in ${\cal N}=4$ SYM.

In~\cite{Gehrmann:2011xn,Ahmed:2016vgl}, it was observed that the highest transcendental part of the QCD form factors match exactly with those of half-BPS and Konishi in ${\cal N}=4$ SYM. In spirit of that we intend to examine if a similar thing happens in our case. So, we compare the finite parts of the form factors, ${\cal F}^{(1)}_{\rm fin}$ and ${\cal F}^{(2)}_{\rm fin}$ in the limit of same virtuality of the singlet operators with the corresponding quantities of di-Higgs boson production through gluon fusion~\cite{Banerjee:2018lfq}. A complete mismatch is found, neither highest nor the lower transcendental terms match to each other.

In the limiting case, when masses of the two massive particles become equal, i.e. $q_1^2=q_2^2=M$, the resulting expression is checked against two different four-point amplitude of double ${\cal O}^{\rm hBPS}$ and ${\cal O}^{\rm nBPS}$~\cite{Ahmed:2019yjt} operators. Comparison with one loop FF of double ${\cal O}^{\rm hBPS}$ reveals almost none of the terms match whereas for double ${\cal O}^{\rm nBPS}$, almost all the highest transcendental terms match exactly but ${\rm Li}_2(-X)\,,\log^2(X)$ and $\pi^2$: 
\begin{align}
    \left[{\cal F}^{(1)}_{\rm fin}-{\cal F}^{(1)}_{{\rm fin,~}{\cal KK}}\right]_{\tau(2)} &= \rho\left( 4 \pi^2+ 12 \log^2(X)-48{\rm Li}_2(-X)\right)\nonumber\\
    {\rm with}~~~\rho &= \frac{ X Y (1+X^2-XY)}{X^4-1}\,.
\end{align}
${\cal F}^{(1)}_{{\rm fin,~}{\cal KK}}$ is the corresponding finite part of the 1-loop FF for two SU(N) scalars to two massive particles of equal masses described by Konishi operator. The subscript $\tau(2)$ implies the transcendental 2 terms. The variable $X$ and $Y$ are defined through $s=M^2(1+X)^2/X$ and $t=-M^2 Y$. At this point it is not fully clear why for certain cases the highest transcendental terms match, we need to explore more cases to reveal the underlying reasons behind the coincidence of this kind.

\section{Conclusions and Outlook}
\label{sec:concl}
In this article, we report the very first calculation of two-loop four-point amplitude involving two SU(N) colored massless scalars and two massive particles with different masses represented through a half-BPS and Konishi operators in ${\cal N}=4$ SYM. Following the Feynman diagrammatic approach and using the state-of-the-art techniques the results of the form factors are presented up to weight 4 terms expressed in terms of $\log\,,{\rm Li}_n$ and ${\rm Li}_{2,2}$ functions. Supersymmetry (SUSY) preserving regularisation scheme, the modified dimensional reduction, is employed in order to regulate the ultraviolet divergences arising from the compositeness of the unprotected operator and infrared singularities resulting from massless particles in the theory. Upon verifying the universal behaviour of the infrared divergences, we show that the results contain not only the highest transcendental terms but also all the lower ones. A detailed comparison, particularly, of the highest transcendental terms with several other four-point amplitudes in ${\cal N}=4$ SYM and QCD is performed which does not provide any decisive pattern. More studies~\cite{Ahmed:2019yjt} of similar kind would be required to shed light on the structures of massive amplitudes.

\section*{Acknowledgements}
We are grateful to S. Seth for providing his python code to incorporate Majorana fermions in \qgraf. We thank V. Ravindran, G. Das, P. Banerjee and A. Chakraborty for various discussions. TA thanks J. Henn for discussion on master integrals, L. Chen for discussions on Kira and M. K. Mandal for discussions on FIRE which helped to execute this project.

\bibliography{n4} 
\bibliographystyle{utphysM}
\end{document}